\documentstyle[11pt,epsfig]{article}
\newcommand{\trr}{\mbox{\rm tr}}
\newcommand{\detr}{\mbox{\rm det}}
\newcommand{\cV}{\mbox{${\cal V}$}}
\newcommand{\tMM}{\mbox{${{}^{\scriptscriptstyle M}}$}}
\newcommand{\tWW}{\mbox{${{}^{\scriptscriptstyle W}}$}}
\newcommand{\suno}{\mbox{${\scriptscriptstyle 1}$}}
\newcommand{\sdos}{\mbox{${\scriptscriptstyle 2}$}}

\newcommand{\spp}{\mbox{${\scriptscriptstyle p}$}}
\newcommand{\sqq}{\mbox{${\scriptscriptstyle q}$}}
\newcommand{\srr}{\mbox{${\scriptscriptstyle r}$}}
\newcommand{\stt}{\mbox{${\scriptscriptstyle s}$}}
\newcommand{\sVuno}{\mbox{$\stackrel{\suno}{\cV}$}}
\newcommand{\sVdos}{\mbox{$\stackrel{\sdos}{\cV}$}}

\newcommand{\spcdots}{\mbox{$\stackrel{\spp}{\cdots}$}}
\newcommand{\sqcdots}{\mbox{$\stackrel{\sqq}{\cdots}$}}
\newcommand{\srcdots}{\mbox{$\stackrel{\srr}{\cdots}$}}
\newcommand{\sscdots}{\mbox{$\stackrel{\stt}{\cdots}$}}

\newcommand{\beq}{\begin{equation}}
\newcommand{\eeq}{\end{equation}}

\newcommand{\FF}{\mbox{${{}_1 F_1}$}}

\columnwidth\textwidth
\begin{document}
\begin{flushright}
BUTP-96/8 \\
\end{flushright}

\begin{center}
\Large{ \bf 
Neutrino Oscillations and the MSW effect 
in Random Solar Matter }
\end{center}

\begin{center}
E. Torrente Lujan.\\
Inst. fur Theoretische Physik, Universitat Bern \\
Sidlerstrasse 5, 3012 Bern, Switzerland. \\[3mm]
e-mail: e.torrente @cern.ch 
\end{center}

\begin{abstract}
We investigate the effects of random density fluctuations on
neutrino oscillations in the Sun environment. We show how
the average of certain quantities which can be used to describe the MSW effect
can be computed analytically.
We examine also the hypothesis commonly accepted that
only perturbations inside the resonance layer can have relevance. 
The average amplitude, which gives the ''coherent probability'', is computed in an
analytical and exact way for the case 
of colored $\delta$-correlated Gaussian noise: the random perturbation induces a renormalization of
the matter density which acquires an imaginary part proportional to the
fluctuation magnitude in the resonance region.
Integral equations are given for the density matrix of the system in the ''optical'' approximation.


\end{abstract}

\newpage

\section{Introduction}

In this work we investigate matter-enhanced neutrino flavor transformations, the
MSW effect \cite{wol1,mik1}, for the case of a exponentially decaying 
matter density with a colored Gaussian noise. 

There are several   examples of interesting
transport and scattering processes induced or modified by the 
presence of a disordered or random medium: dislocations in crystals and the
solid-liquid transition; random impurity potentials
produce the localization of
quantum wavefunctions which enables one to understand the transition between
insulators and conductors; resistance anomalies at
 low temperature and in presence of magnetic fields of ''weak localized'' electron
systems subject to a random potential  (\cite{ber1,vol1}).
The same basic localization phenomena
explain also in optics the backscattering
enhancement in electromagnetic wave scattering from a randomly rough surface
(\cite{fitz1}). Usually, a strong
dependence on the dimensionality of the problem is observed.
In some other cases as in nuclear physics the consideration of random matrix
hamiltonians allows the simplification of otherwise unmanageable systems.

For neutrino oscillations in presence of random
 or rapidly twisting
magnetic fields considerable  work
has been done already \cite{lor1,ane1,akh1,nic2}.
Random density fluctuations have received some attention only recently.

In \cite{lor1}, a differential equation for the averaged survival probability
was derived for the case in which the random noise was taken
to be a delta-correlated white Gaussian distribution.
The differential equation was solved numerically and the neutrino evolution obtained.
Arguments were given which indicate that if the correlation of the matter
density fluctuations is small compared to the neutrino oscillation length at
resonance, one obtains the same result as for the case of a delta-correlated
noise.
In \cite{lor2} the more realistic case of colored noise 
was considered also in a numerical way
and applied to Supernova dynamics. An approximate differential equation for the
averaged survival probabilities was obtained using the 
 hypothesis that the fluctuations should not affect the evolution far from the
resonance.
In \cite{nun2} ,
the implications of random perturbations 
upon the Solar neutrino deficit are considered numerically. It is found that
 the MSW effect is
 rather stable under these fluctuations
specially in the small mixing case but in anycase the experimental
$(\Delta m^2,\cos\theta)$ exclusion curves get modified in an appreciable way.

In this work we try to develop analytical results 
for neutrino flavor
oscillations induced by a random matter density. 
We are interested mainly in the persistence of the MSW effect. The obtention of
concrete values for the survival probabilities and phenomenological consequences
for the solar neutrino problem is left for a subsequent work.
The obtention of these phenomenological consequences is
not an easy task , at least, because
the amplitude of possible 
density fluctuations in the Sun is poorly known 
experimentally and theoretically. Different arguments can give easily values for it differing by two orders
of magnitude (\cite{nun2}): between $0.1\%-10\%$ of the local density. Stronger local inhomogeneities,
for example at the near-surface  dark spots should not be discarded.

Our basic starting point will be the
exact analytical solution obtained in \cite{emi1} for the
 neutrino oscillation
amplitudes
 in presence of an exponentially decaying matter density. The random component
is consider as a perturbation to this solution.
Inspired by the electromagnetic wave scattering in random media,
{\em coherent } and {\em incoherent} transition probabilities are defined. The
basic result of this work is that the coherent probability, which comes
essentially from an averaged amplitude, can be computed exactly in some cases.
Different integral equations are given for the incoherent probability using an
''optical'' approximation. Approximate solutions valid in a limited range are
obtained for them. 
By the very nature of the procedure the results are valid or easi    ly generalizable
to any number of neutrino species.

The outline of this work is as follows. The first section, after the
introduction of some known results, is dedicated to
explore diverse quantities whose  stochastic average can be computed exactly
or at least by an easy approximation.
One of these quantities is the determinant of the 
evolution operator of the system. We use a ''naive'' argument to show the
plausibility that the influence of the random perturbation on the neutrino
oscillations can be described by a complex 
redefinition of the matter density. 
This suggestion allows to define some {\em ansatz} probability.
Another quantity is the ''total cross
section'' of the system, we suggest the interest in further studying this
quantity which can contain information on the presence and localization of the
MSW resonance.  
In the same section, 
the physical supposition that only those random perturbations
happening in the resonance layer can have importance  is studied briefly . We argue that this
supposition must be taken with care, because, as we show, even small phase
shifts {\em before} the resonance region can have 
some appreciable importance in the final survival probability.

In the next section we define some perturbative expansion for the density
matrix of the system. 
The {\em coherent} and {\em incoherent} parts of this matrix
 are defined. The coherent part being in some sense
the zero-order approximation for the full density.
Using an ''optical'' approximation, that is discarding a certain class of terms
in the perturbative expansion, a very general 
 integral equation for the averaged density is
derived. In this integral equation, basic ingredients are
both  the coherent density and the averaged
amplitude and the two point correlation of the matter density perturbation.

In Section (\ref{s4}) different particular cases are considered. A simpler
expression for the previous integral equation is given for the case where the
random perturbation is $\delta$-correlated. A further simpler expression is
given for the small mixing case of two neutrino species. In this case upper
limits for the total averaged probability can be obtained depending on the
coherent probability and an autocorrelation integral.

In Section (\ref{s5}) the coherent part is computed. In a first case the small
mixing condition and
different approximations are used. Averaging the amplitude 
amounts to the multiplication by a certain slowly time-varying diagonal matrix.
In a second particular case, it is shown how  the average  can be computed
exactly. The effect of the averaging is indeed 
a complex renormalization of the initial
matter density as it was  suggested earlier. 
Even if it is a very particular case, it is shown how it can have relevance in
more realistic computations.
Finally, making use of these averages, coherent survival probabilities and the
''cross-sections'' previously defined are computed.

\section{Preliminary Considerations}
\label{s2}
\subsection{The non-stochastic solution}

The solution for the neutrino oscillations in solar matter 
 described by the equation
\beq
i\partial_t \nu = (H_0+\rho_0 \exp (-\lambda t) u A u^{-1}) \nu
\label{e3001}
\eeq
has been given in \cite{emi1}.

 For $t\to\infty$, the solution can be written as ($\lambda$ will be understood
generally set to 1)
\beq
\nu(t)=\exp -i H_0 t\ \  U_{r}(\rho_0)\ \nu(0)
\label{e1001}
\eeq
For any arbitrary time 
\beq
\nu(t)=U(t,t_0) \nu(0)
\label{e1051}
\eeq
with
\beq
U(t,t_0)=U_s(t)^\dagger U_s(t_0),\quad\quad U_{s}(t)=
U_{r}\left (\rho_0 e^{-t}\right )\; \exp-i H_0t 
\label{e6003}
\eeq
The solution can be extended to complex $\rho_0$, then $U_s$ is not
unitary but still keeps the same functional form, the
 previous expression must be
changed (following the same argument used to derive Eq.(58) in \cite{emi1}) to 
$$U(t,t_0)=U_s(t)^{-1} U_s(t_0)$$

In the general case ($\rho_0$ real or complex)  the coefficients of $U_{r}$ in
the Expression (\ref{e1001}) are just confluent generalize
hypergeometric functions of one order less than the dimension of the problem, with argument
$-i\rho_0/\lambda$ and parameters which are combinations of the eigenvalues of
$H_0$ and the elements of the mixing matrix u.
For $\rho_0$ real, $U_{r}$ becomes unitary and its coefficients adopt a form
particularly simple and symmetric.
As example, we write it explicitly for the two-dimensional case:
\begin{eqnarray}
U_{r}(\rho_0)&=&\pmatrix{
 F &  \frac{V_{12}}{V_{11}}  \  G \cr
 -\frac{V_{21}}{V_{11}}
\ \exp(z) \ G^\ast& \ \exp z \ F^\ast \cr }
\label{e150}
\end{eqnarray}
with the shorthands
\begin{eqnarray}
G=g(z)= \frac{V_{11}}{1+\beta} \FF (1+V_{11}\ \beta, 2+\beta;z) 
,\quad F=\FF (V_{11}\ \beta, \beta;z), \nonumber\\
\nonumber\\
 z=-i \rho_0, \quad \beta=\Delta m^2/2 E  \hspace{2cm}\nonumber
\end{eqnarray}

The algebraic properties of F and G guarantee automatically the
unitarity of $U_{r}$. Its determinant is
\beq
\detr\ U_{r}(\rho_0)=\exp -i\rho_0
\eeq

The survival probability is given by
\begin{eqnarray}
P_{ee}(\rho_0,\beta,\theta)=&1-2 S^2\theta\left (1+ C^2\theta \left \|
\FF\left ( i\beta C^2\theta, 1+i\beta; 
-i\rho_0 \right ) \right\|^2\right ) \label{e7006}
\end{eqnarray}
or in the small mixing angle limit:
\begin{eqnarray}
P_{ee}(\rho_0,\beta,\theta)=& \left \|
\FF\left ( i\beta C^2\theta, i\beta; 
-i\rho_0  \right ) \right\|^2
\label{e7007}
\end{eqnarray}

\subsection{The average determinant}
\label{s22}
In this work we are interested  to investigate 
what is the effect of introducing
a random density perturbation, that means, 
when $\rho(t)=\rho_0\exp(-\lambda t)$
is changed to $\rho(t) (1+\delta (t))$ where $\delta$ is a stochastic Gaussian
process of zero mean, characterized completely by a certain two point correlation function.

Having in mind the optical potentials 
and random matrix approximations used in nuclear
physics \cite{aga1} and wave scattering in random media \cite{fitz1}, one could think that in our case the effect of such
introduction should be approximately equivalent to a redefinition of the function
$\rho(t)$ which in the most simple case would amount just to a
renormalization of the constant $\rho_0$ (to a complex value in general in
analogy with the complex wave numbers appearing in random media wave scattering).

We get some hint that such supposition is reasonable if we consider the
average value  of the determinant of the evolution operator of our differential
equation. To compute that determinant and its average value is a trivial task.

The evolution operator  can be expressed formally 
as a time-ordered (T) integral:

\beq
U(t,t_0)=T\ \exp -i\int_{t0}^t ds H(s)
\label{e102}
\eeq

 Its determinant is simply the elementary exponential:
\beq
\detr\ U= \exp-i\int_{t0}^t ds {\rm tr}\ H(s)
\eeq

As $H(t)$ is assumed to be a Gaussian process, the statistical average is:
\beq
<\detr\ U>= < \exp-i\int \trr\ H> = \exp -i\int <{\rm tr}
H>-\frac{1}{2}\int\int <<{\rm tr} H(t){\rm tr} H(s)>>
\eeq

Where $<<A\ B>>=<A\ B>-<A><B>$.\footnote{We will use indistinctly the notations
$<A>,\ \overline{A},\ $ or $ {\cal A}$ for the average of the variable A}

So, for the total hamiltonian H defined in Eq.(\ref{e3001}) we have: 
\beq
<\trr\ H(t)>= \Sigma +\rho_0 \exp -t
\eeq
The constant term $\Sigma$ is unimportant, it can be set to zero by a convenient
redefinition of the zero energy.
With
\beq
<\delta(t)\delta(s)>=k(t+s) g(\mid t-s \mid)
\eeq
we have
\beq
<<\trr H(t)\ \trr H(s  )>>=k(t+s)\rho_0^2\exp-(t+s )\
g(\mid t-s \mid)
\eeq

We are interested mainly in the limit $t\to\infty$, we take also  $t_0=0$. After some elementary integrations we arrive at:
\beq
<\detr\ U>= \exp -i\int_0^\infty dt\  \rho_0
\left(1-\frac{1}{2}\rho_0 i k(t)\int_0^t g(u)du\right )\exp( -t)
\eeq

We can suppose with generality
that $g(u)$ goes to zero quickly for $u\to\infty$
, that  guarantees that we make a
relatively small error if we substitute the upper limit of integration by $\infty$. 
For the case of $g(u)$ being a delta function, this approximation becomes exact.
We will suppose also that the factor $k(t)\approx k $, constant.

So we can interpret that 
the introduction of the random perturbation
 induces a renormalization of the initial density which is  proportional
to the integral of the autocorrelation function.
\beq
\rho_0\to \rho_{0r}=\rho_0\left (1-i\frac{k\rho_0}{2}\int_0^\infty g(u)du\right )
\eeq

Under this approximation, the average determinant becomes:
\beq
<\detr\ U>=\exp -i\rho_{0r}
\eeq 
the  renormalization introduces an imaginary part that renders
the operator $U(\rho_{or})$ non-unitary, on average;
certainly
the initial hamiltonian would become non hermitic 
if we would substitute
directly it it $\rho_{o}$ by $\rho_{or}$.
Formally we can recover at least the condition
$ \mid <\det U>\mid =1 $ (but of course not the unitarity condition)
adding to the initial hamiltonian a diagonal term of the form
$$H_r=i\rho_0^2\exp-t\int_0^\infty du g(u)\ \ I$$
I being the identity matrix. This amounts to a (complex) shift in the energy 
which  is unobservable.

We note that for computing Eq.(\ref{e7006}), 
the unitarity of U 
has been explicitly used in an
important way 
(\cite{emi1}).
Our temptative {\em ansatz} is to suppose that the introduction of a
 stochastic term is equivalent to the consideration of a non-stochastic equation with a
redefined initial maximal density, as obtained before, plus appropriated
''counterterms'' that  render the full hamiltonian hermitic.
 The physical
information, the transition probabilities, can be computed by analytical
continuation of the original transition probabilities corresponding to the
initial equation with the new redefined parameters.

So under this ansatz  the averaged transition
probability in presence of the random term is (for two dimensions, Eq.(\ref{e7006}))
$P_{ee}(\rho_{or})$.

The rest of this article will be devoted essentially to a more rigorous
justification of this prescription. We will show that
at least for the, so called, {\em coherent} probability this assumption is true
in a particular case.

\subsection{ Cross sections and the  Optical Theorem.}

As we will see later, while it is relatively easy to compute the average values
$<U_{ij}>$, the averaged probabilities, which depend on quadratic quantities $<\mid
U_{ij}\mid^2>$, are nearly inaccesibles analytically.

In fact, apart from survival probabilities, we are also
simply interested 
to study
whether the MSW effect survives or to which extent gets
modified with the introduction of a random perturbation. 
It would be important  to find
a quantity that both: gives us some information about
the existence, amplitude or position in parameter space of the MSW effect 
and its stochastic average is easy to compute .
In this sense, we propose to use the ''scattering'' matrix T defined by
\beq
U_{I}(t\to\infty)=1+T
\eeq
where $U_I$ is the  evolution matrix defined before in some appropriate interaction
representation.

T satisfies an ''Optical Theorem'':
\begin{eqnarray}
U_I U_I^\dagger=1&=&(1+T)(1+T^\dagger) \\
     T+T^\dagger&=&-TT^\dagger
\end{eqnarray}

In particular we define the two quantities:
\begin{eqnarray}
\sigma_1^i&=&-2\Re T_{ii}=\sum_k \mid T_{ik}\mid^2\\
\sigma_2&=& -2 \Re \trr T= \sum_{ik}\mid T_{ik}\mid^2\equiv \mid\mid T\mid\mid^2
\end{eqnarray}

We can identify the first expression with a total cross section in particle
physics (sum over all final channels). The second is the sum of all total cross
sections over initial channels. We expect that the quantities
$\sigma_1,\sigma_2$ can give us interesting information about
 the scattering
process, hopefully the MSW resonance should manifest itself in them.
The important thing is that, because both depend linearly on T (U), their statistical average is rather easy to
compute.

In the mass basis
\begin{eqnarray}
\tMM\sigma_1^i=-2\Re (U_{ii}-1)= 2(1-\Re U_{ii})
\end{eqnarray}
In the weak basis $U_I=uU_r u^{-1}$, and we have a similar expression. 
The expression for $\sigma_1^1$  is particularly simple 
\begin{eqnarray}
\tWW\sigma_1^1=-2\Re ((uU_r u^{-1})_{ii}-1)= 2(1-\Re \trr U_r V)
\label{e6024}
\end{eqnarray}
On the other hand, $\sigma_2$ is basis invariant and (d is the dimension of the problem)
\begin{eqnarray}
\sigma_2=2(d-\Re\trr U_r)
\label{e6025}
\end{eqnarray}

\begin{figure}[p]
\centering\hspace{0.6cm}
{\epsfig{file=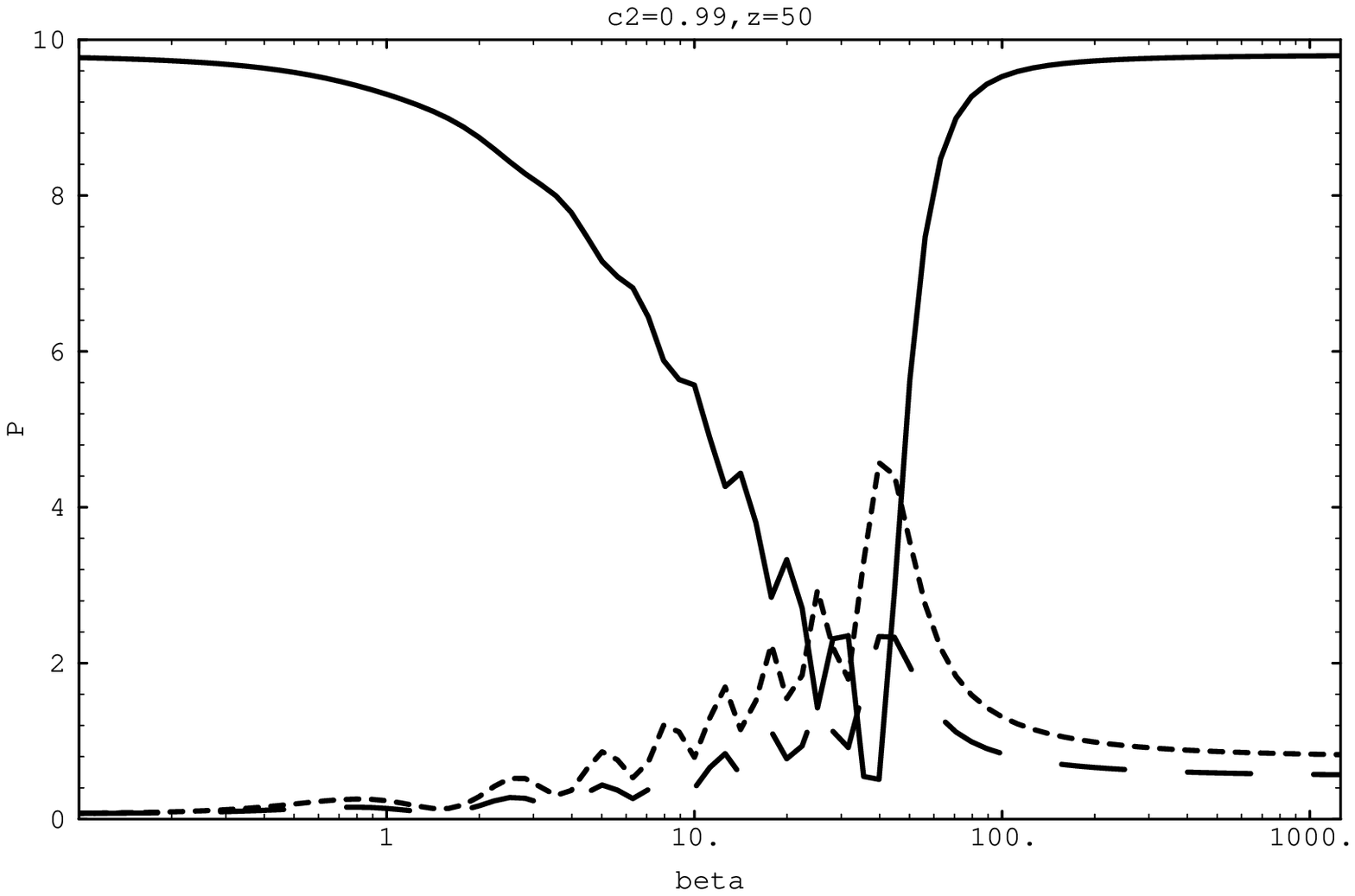,height=12cm}}
\caption{The ''total cross sections'' 
\protect{$\tWW\sigma_1^1$} (short dash) and 
\protect{$\sigma_2$} (longer dash)  for a neutrino 
produced at the Sun (\protect{$\rho_0/\lambda=50,\cos^2=0.99$}) as a function of \protect{$\beta=\Delta m^2/2E\lambda$}
 (see Eqs.(\protect\ref{e6024}-\protect\ref{e6025})). 
The continuos line is the survival probability (\protect{$\times 10$}).} 
\label{f1}
\end{figure}

In Fig.(\ref{f1}) we plot  the quantities $\tWW\sigma^1_1,\sigma_2$ for
particular parameters together with the
$\nu_e$ survival probability. We see that both  show a 
prominent peak in the
resonance region. They reproduce 
the secondary extrema as well .

For $\beta\to 0$ or $\beta\to\infty$, $\sigma_2\approx 0$ this implies
\beq
2\approx\Re\trr U_r\equiv (1+\cos\rho_0)\Re F+\sin\rho_0 \Im F
\eeq
For the value of $\rho_0$ used in the plot: $1+\cos\rho_0\approx 2,
\sin\rho_0\approx 1/5$ (a small but finite value). We know also that $P_{ee}\simeq\mid
F\mid^2\leq 1$, then $\Re F\sim 1$ and $P_{ee}\sim 1$.

On the other hand, for $\beta\to\beta_{res}$, $\sigma_2= 4,\tWW\sigma_1^1=2$.
$\Re\trr U_r=0, \Re\trr U_r V=0$ imply
$\Im F, \Re F\approx 0$ and we obtain the resonance 
probability $P_{ee}\approx 0$.  

 Later we will show the same type of plot for the averaged quantities.

\subsection{The influence of the resonance layer}

From a physical point of view, and, as is usually assumed, the only
perturbations that can have an influence on the final transition probabilities
are those which happen inside the resonance layer.

We have supposed Gaussian random perturbations; there is a finite probability
that in any moment through the neutrino path, a strong local fluctuation makes
the density term similar to the difference $\Delta E$ provoking in this way a 
resonant level crossing. Numerical studies (\cite{nun2,lor1,lor2}) seem to show
that this possibility in fact doesn't happen easily
 at least for small
mixing angles.

But, we want to show that 
small  changes of the neutrino wave function {\em before} arriving to the
resonance region can have an important effect on the final probability.

Let's suppose that for any circumstance the flavor components of the neutrino wave
function acquires a relative phase $\phi$ at some $t_1$ much later than its creation time.
At infinite its wave function would be
\begin{eqnarray}
\nu(t\to\infty)&=&U(t\to\infty,t_1)\pmatrix{1 & 0\cr 0 & e^{-i\phi}
}U(t_1,t_0)\nu_0 \nonumber \\
&=& U_s^\dagger(\infty)\ \Sigma\ U_s(t_0) \nu_0
\end{eqnarray}
Where the matrix $\Sigma$ is defined by the above equation.

In Section~(\ref{s502}) we will see that, although the similarity is not
complete, indeed the effect of random perturbations can be accounted for by 
the insertion of a certain  matrix inside the non-random U.

\begin{figure}[p]
\centering\hspace{0.6cm}
\begin{tabular}{c}
{\epsfig{file=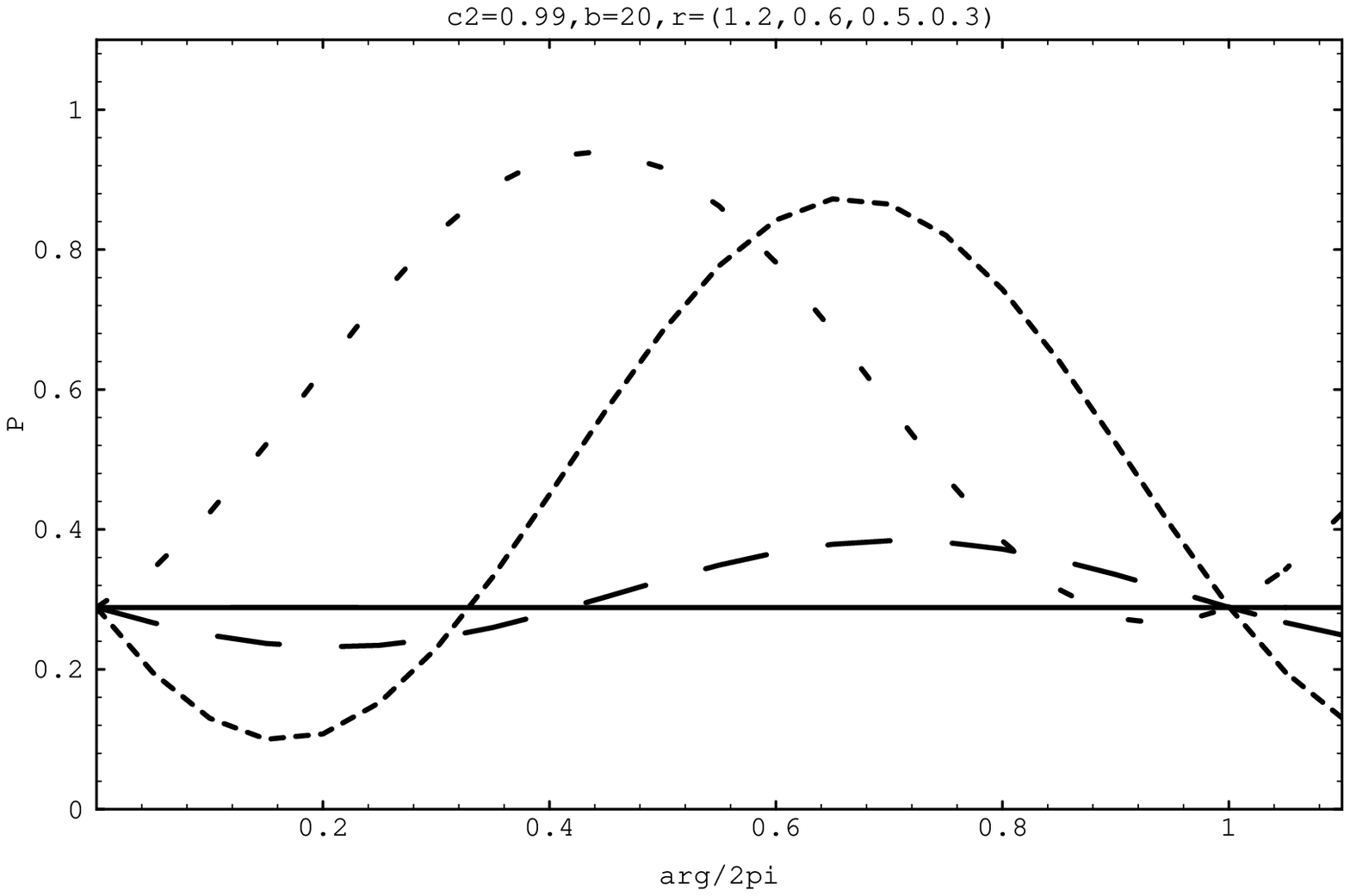,height=8cm}}\\[-2cm]
{\epsfig{file=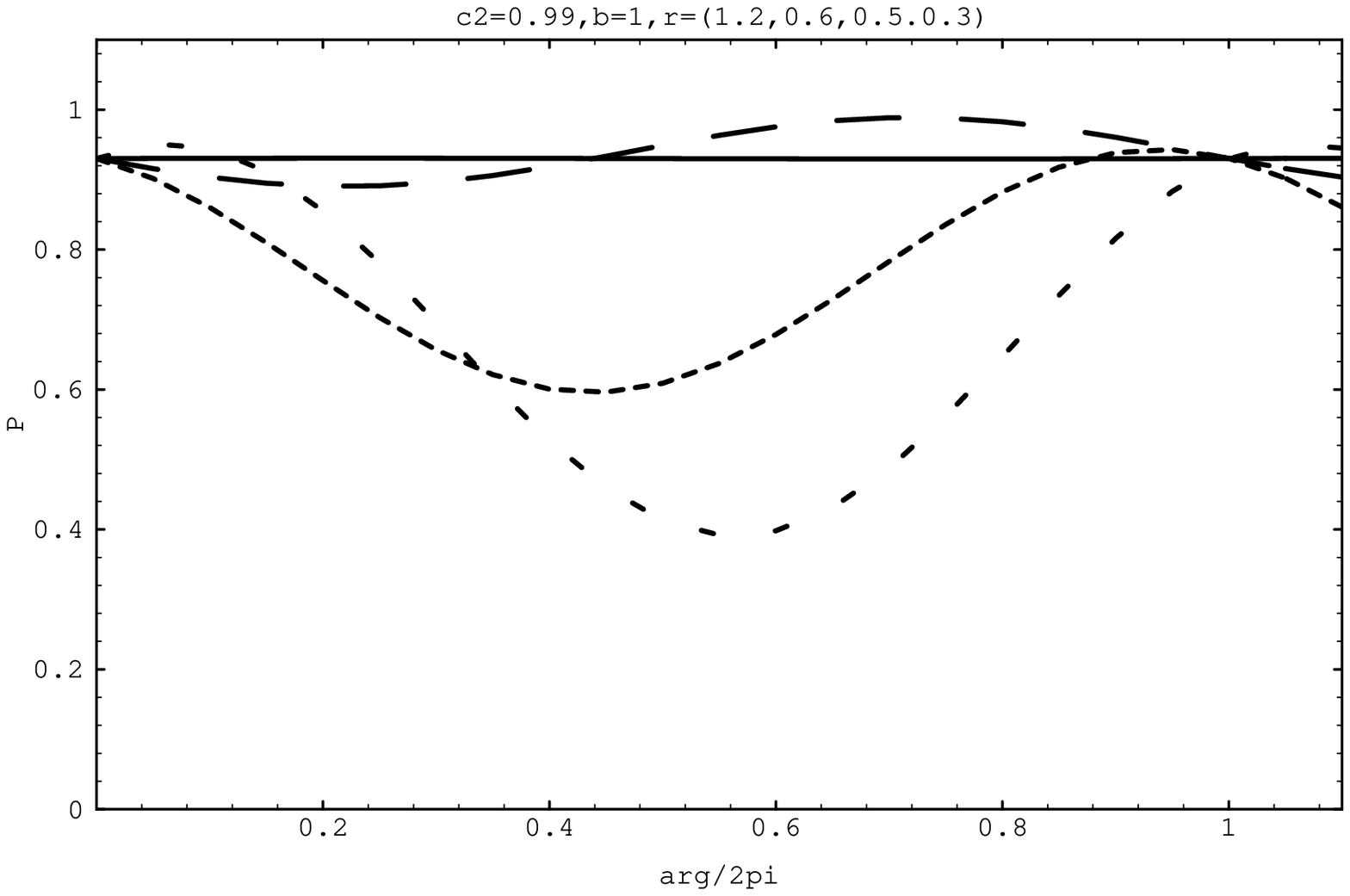,height=8cm}}\vspace{-1cm}
\end{tabular}
\caption{The survival probability as a function of an arbitrary phase shift
introduced in different positions along the trajectory
of a neutrino created near the Sun core 
 ( $\cos^2\theta=0.99$,$\beta=1,20$) .
For $\beta=20$ the resonance is situated at $r/r_0\approx 0.5-0.6$.  
The continuos line correspond to 
$r/r_0=1.2$: well ahead the
resonance region. The longer dashed to $r/r_0=0.6$: inside the resonance. The
largest variation (both shorter dashed lines) correspond to
$r/r_0=0.5,0.3$ at the beginning or clearly before the resonance respectively.}
\label{f2}
\end{figure}

In Fig.($\ref{f2}$) we show the survival probability as a function of the
intermediate time $t_1$ (or $r/r_0$ in the figure)  and $\phi$.
We see that the phase shifts introduced after the resonance layer have no  or 
very little effect. The shifts introduced at the beginning or very clearly
before  the resonance region can have a drastic influence on the final
probability.
The obvious conclusion  from this is that if random perturbations affect
(even slightly)
 the phase of the wave function long before the resonance region,
then they  can have an appreciable effect on the final
survival probability.

\section{Formulation of the Main Approach.}
\label{s3}

The  density operator of any system defined by a hamiltonian
$H(t)=H_0+W(t)$ and  certain initial conditions, is given, in terms of U, the
evolution operator, by:
\beq
\rho(t)=U(t,t_0)\rho(t_0) U(t,t_0)^\dagger
\eeq

If the potential
vanishes for $t\to\infty$, the  asymptotic  propagator obeys a free
Schroedinger equation and can be cast in the form:

\beq
\rho(t\to\infty)=\exp -i H_0 t\ \rho_{as}\ \exp i H_0 t 
\label{e4029}
\eeq
$\rho_{as}$ is a time-independent operator to be determined  for any particular
problem and initial conditions.

Our objective will be to derive an integral equation for the statistical
average of the density operator 
$\overline{\rho(t)}$
in the $t\to\infty$ limit
for  the special type of stochastic potential
we have discussed in the previous section.

Using perturbation theory around the known solution of the non-stochastic part
of $W(t)$, $V^0(t)\equiv<W(t)>=\rho_0\exp(-t) V$, $V^2=V$ we have for U:

\beq
U(t,t_0)=U^0(t,t_0)+\sum_n U^{(n)}(t,t_0)
\eeq

As we have seen before,
for this kind of potential 
we can factorize the operator $U^0$
in ''creation'' and ''anhilation'' operators:
\beq
U^0(t,t_0)=U_s^{0\dagger}(t)U_s^0(t_0)
\eeq
so
\begin{eqnarray}
U(t,t_0)&=&U_s^{0\dagger}(t)\Sigma(t,t_0) U_s^0(t_0) \\\nonumber \\
\Sigma(t,t_0)&=& 1+\sum_{n=1}^\infty \int_{t0}^t d\tau \ V_I(\tau_n)\dots
V_I(\tau_1) \\
(d\tau &=& d\tau_1\times \dots \times d\tau_n) \nonumber \\
\end{eqnarray}
where
\beq
V_I(t)=  U_s^0(t) \left (-iV(t)\right ) U_s^{0\dagger}(t),\quad
V(t)=W(t)-V^0(t)
\label{e2051}
\eeq
We define equally
 the density matrix in the interaction representation
\begin{eqnarray}
\rho_{I}(t)=U_s^0\rho U_s^{0\dagger}=\Sigma(t,t_0)\rho_{I}(t_0) \Sigma(t,t_0)^\dagger
\label{e4035}
\end{eqnarray}

In the limit $t\to\infty$, $U_s^0(t)\to \exp i H_0 t$
and we identify
\beq
\rho_{as}=\rho_{I}(t\to\infty)=
\Sigma(\infty,t_0)\  \rho_{I}(t_0)\  \Sigma(\infty,t_0)^\dagger
\label{e4037}
\eeq

The average density operator contain a contribution from neutrinos which have
scattered coherently (or specularly in the electromagnetic wave analogy). This 
contribution is  obtained averaging the amplitude U:
\beq
<\rho^{coh}>= <U>\rho_0 <U^\dagger>
\label{e7037}
\eeq

The contribution from the incoherent, or diffuse, component is obtained by subtraction.
\beq
<\rho^{incoh}>= <\rho>-<\rho^{coh}>
\eeq

All the ''randomness'' information is  included
in the operator $\Sigma$, so the Eq.(\ref{e4035}) 
holds also for the respective  statistical 
averages: $<\rho>_I=<\rho_I>$.

Taking\footnote{There  
should not be risk of confusion between 
 $\rho,\rho_0$ as
density matrices or matter densities.} 
for the initial density matrix
$$\rho_0=u(1,0,\dots,0)u^{-1}=V$$
the macroscopical $\nu_e\nu_e$ transition probability is, 
\begin{eqnarray}
P_{ee}^M(t)=\trr <\rho(t)> V
\label{e4040}
\end{eqnarray}

Taking $t\to\infty$ and averaging out oscillatory terms
\begin{eqnarray}
P_{ee}^M=\Sigma_{i} \rho_{as,ii} V_{ii}\quad ( \simeq \rho_{as,11}\; \hbox{\rm
for small mixing: $V_{11}\to 1$})
\label{e4041}
\end{eqnarray}

From Eq.(\ref{e4035}),
the statistical average of $\rho_I$ can be decomposed in a sum of terms of the form:
\begin{eqnarray}
\overline{\rho_I}&=&\sum_{p,q} A_{pq} \nonumber\\
A_{pq}&=&<\int V_I \stackrel{p}{\cdots} V_I\ \rho_{0I}\ \int V_I^\dagger \stackrel{q}{\cdots} V_I^\dagger>
\end{eqnarray}

It is assumed that the stochastic potential is a Gaussian 
process with zero mean.
This implies that the average of terms with an odd number of V's is zero. On the other hand the average of the terms 
with a even number of V's can be decomposed in a sum
of two-V averages extended over all possible combinations.
For example
\beq
\overline{VVVV}= \overline{VV}\ \overline{VV}+ \stackrel{\suno}{\cV}\overline{VV}\stackrel{\suno}{\cV} +
\stackrel{\suno}{\cV}\  
\stackrel{\sdos}{\cV}\  
\stackrel{\suno}{\cV}\  
\stackrel{\sdos}{\cV}
\label{e1002}
\eeq
The numbers over the {\em calligraphic V's} identifies the pairs which are averaged.  

According to arguments developed in \cite{aga1,lor1} the terms which
contain "cross" averages, for example the last term in Expression
 (\ref{e1002}),
can be discarded to a good approximation for pair correlation functions which are
significantly non-zero only for a relatively short time difference (for example for delta-functions). We will
suppose that this is always true in our case;  in the particular case of
the coherent part we will not need this approximation at all.

For convenience we will use from now on
a ''two time'' density matrix,
the statistical average of $\rho_{I}$ is given by
\begin{eqnarray}
\overline{\rho_{I}(t,s)}&=&\overline{\rho_I^{coh}}+\nonumber\\
&&+\sum_{p,q,r,s}\int\overline{V\spcdots V}\sVuno\overline{V\sqcdots V} \rho_{0I} \int\overline{V\srcdots
V}^\dagger \sVuno^\dagger \overline{V\sscdots V}^\dagger+  \nonumber\\
&&{\vphantom{\sum_r \int}}
+\sum\int\overline{V\dots V}\sVuno
\overline{V\dots V}\sVdos \overline{V\dots V} \rho_{0I} \int\overline{V\dots
V}^\dagger \sVdos^\dagger
 \overline{V\cdots V}^\dagger \sVuno^\dagger  \overline{V\cdots V}^\dagger 
+\nonumber \\ 
& &{\vphantom{\sum_r \int}}+\; \cdots  
\label{e1025}
\end{eqnarray}

The first summand is the coherent part
and  its computation amounts to sum  the series
\beq
<1+\int V_I+\int V_I \int V_I+\dots>=<T\exp \int V_I >\equiv \overline{\Sigma}
\label{e1061}
\eeq
Note that the average can be performed exactly without neglecting "crossed"pair
averages, so the last relation Eq.(\ref{e1061}) can be considered exact.
 Interchanging averaging and time ordering:
\beq
<T\exp \int_{t0}^t V_I >=T <\exp \int_{t0}^t V_I>= T \exp \frac{1}{2}\int_{t0}^t\int_{t0}^t <<V_I(s_1)V_I(s_2)>>ds_1ds_2
\label{e1026}
\eeq

From this expression an effective potential $V_I^{eff}$ can be defined which summarizes the
influence of the stochastic potential and such that
\begin{eqnarray}
\Sigma^{eff}&\equiv& T\exp \int_{t0}^t V_I^{eff}(t')dt' = \overline{\Sigma}\label{e7465}\\[0.2cm]
V^{eff}_I(t')&\equiv& \frac{1}{2}\int_{t0}^t <<V_I(t')V_I(s)>>ds 
\label{e1006}
\end{eqnarray}

We can derive an alternative approximate expression ignoring crossing terms in Eq.(\ref{e1061}). 
The propagator computed in such a case will be called the optical propagator in analogy with nuclear physics. It is easy to check
that the optical propagator satisfies the equations
\beq
\Sigma^{opt}(t,t_0)=1+\int_{t0}^t d\tau \ V_I^{opt}(\tau)
\Sigma^{opt}(\tau,t_0) 
\label{e1027}
\eeq

\beq
V^{opt}(\tau)=\int_{t0}^{t }d\tau_2\  
\sVuno_I(\tau)
\Sigma^{opt}(t,\tau_2) \sVuno_I(\tau_2)
\label{e1028}
\eeq

In general $V_I^{opt},\Sigma^{opt}$ are not neccesarily the same as the
$V_I^{eff},\Sigma^{eff}$ defined previously. Also it is not neccesarily more
difficult to compute the former than the latter. In the important
particular case of a
$\delta$-correlated potential both coincide.

We can compute all the further terms 
 in the expansion of $\overline{\rho_{I}}$ (Eq.(\ref{e1025}))
making use of the following expression; 
for an arbitrary non-stochastic operator K, the sum of the series:
\beq
S(t,s)=<K(t,s)+\int_{t0}^t V(\tau) K(\tau,s)+\int_{t0}^t V(\tau_1) \int_{t0}^{\tau1} V(\tau_2) K(\tau_2,s)+\dots>
\eeq
is equivalent to solve the integral equation
\begin{eqnarray}
S(t,s)=K(t,s)+\int_{t0}^t d\tau\ V(\tau) S(\tau,s).
\end{eqnarray}
and to apply the statistical average.
The solution is given by
\beq
S(t,s)=< T\exp \int^t_{t0} V\times\left [ K(0,s)+ \int^t_{t0}d\tau\ 
\left (T\exp \int^\tau_{t0} V\right )^{-1} > \partial_\tau K(\tau,s)\right ] 
\eeq
which, using the properties of the time ordered exponential and putting
$K(0,s)=0$, is equal to
\beq
S(t,t_0)= \int^t_{t0}d\tau <T\exp \int^t_\tau
V > \partial_\tau K(\tau,s) 
\eeq

Using this formula iteratively in Eq.(\ref{e1025}) we obtain
 the expression:
\begin{eqnarray}
\overline{\rho_{I}^{incoh}(t,s)}&=&
\int_{t0}^t\Sigma^{eff}(t,\tau) \sVuno_I(\tau) \Sigma^{eff}(\tau,t_0) 
\rho_{0I} \left (\int_{t0}^s \cdots \right )^\dagger \nonumber\\
&&+\int_{t0}^t \Sigma^{eff}(t,\tau)\sVdos_I(\tau) \int_{t0}^\tau\Sigma^{eff}(\tau,\tau_2) \sVuno_I(\tau_2) \Sigma^{eff}(\tau_2,t_0) 
\rho_{0I} \left (\int_{t0}^s \cdots\right )^\dagger \nonumber\\
&&{\vphantom{\int\sVuno^\dagger}} + \; \cdots 
\label{e1007}
\end{eqnarray}

The operators to the right sides of $\rho_{0I}$ are equal to 
those in the left sides, changing the limit of integration and taking 
the hermitic conjugate.

As it can be seen by explicit developing,
the series given by Eq.(\ref{e1007}) is equivalent to the following
integral equation for the total density $\rho_I$:
\beq
\overline{\rho_{I}(t,s)}=
\Sigma^{eff}(t,t_0)\rho_{0I}\Sigma^{eff\dagger}(s,t_0)
+\int_{t0}^t\int_{t0}^s\Sigma^{eff}(t,\tau_1)  \sVuno_{I}(\tau_1)\ \overline{\rho_{I}(\tau_1,\tau_2})\ \sVuno_I^\dagger(\tau_2)
\Sigma^{eff\dagger}(s,\tau_2)
\label{e1029}  
\eeq
and, at least formally, we have solved the problem in the approximation which
neglects ''crossed'' terms.

\section{The Integral Equation for $\delta$-correlated noise.}
\label{s4}

Let's suppose that the two point correlation matrix has the form:
\beq
<V_I(t) V_I(s)>= -i k(t)\rho(t) \delta (t-s) V_I (t) 
\label{e4056}
\eeq
or more generally, for any non-stochastic operator $K(t,s)$,
\beq
<V_I(t)\ K(t,s)\ V_I(s)>(\equiv \sVuno K \sVuno) =  k(t) \delta (t-s) V_I(t)\ K(t,t)\ V_I(t) 
\label{e1036}
\eeq
The integral equation for the
operator density becomes  (Eq.(\ref{e1029}))
\beq
\overline{\rho_{I}(t)}= \Sigma^{eff}(t,t_0)
\rho_{0I}\Sigma^{eff\dagger}(t,t_0)+\int_{t0}^t d\tau k(\tau)\Sigma^{eff}(t,\tau)V_I(\tau)\overline{\rho_I(\tau)}V_I^\dagger(\tau)\Sigma^{eff\dagger}(t,\tau)
\eeq
Or recalling the definition for
the representation ''I'',  Eqs.(\ref{e2051}-\ref{e4035}): 
\beq
\overline{\rho(t)}=\overline{\rho^{coh}(t)}+\int_{t0}^t d\tau\ k(\tau)\
U(t,\tau)V(\tau)\overline{\rho(\tau)} V^\dagger(\tau)U^\dagger(t,\tau)
\eeq

The probability $P_{ee}^M$ of  $\nu_e$ survival is computed using
Eqs.(\ref{e4040}-\ref{e4041}).
We have the  expression
\beq
P_{ee}^M(t)=P_{ee}^{coh}(t)+\int_0^t d\tau\ k(\tau)\rho^2(\tau)\ \trr
U(t,\tau)V\overline{\rho(\tau)}V U^\dagger(t,\tau) V
\label{e1020}
\eeq
Using the property $V^2=V$, this  expression  can be written also in the form
\beq
P_{ee}^M(t)=P_{ee}^{coh}(t)+\int_0^t d\tau\ k(\tau)\rho(\tau)^2\ \trr
U_R(t,\tau)\overline{\rho_R(\tau)}U^{\dagger}_R(t,\tau)
\label{e5061} 
\eeq
Where $X_R=V X V$.
At this point further approximations are neccesary in order 
to simplify the expressions in eqs.(\ref{e1020}-\ref{e5061})..

\subsection{The small mixing approximation.}
 
In the small mixing, two dimensional case, i.e. taking the limit
\beq
V,\rho_0\to\pmatrix{1&0\cr 0 & 0}
\eeq
such that $X_R=VXV\to X_{11}\rho_0\; $ 
the integral which appears in Eqs.(\ref{e1020}-\ref{e5061})
becomes 
\beq
\int_0^t \rho^2(\tau) k(\tau) \mid U_{11}(t,\tau)\mid^2 \overline{\rho_{11}(\tau)}
\eeq

In this limit, the weak and mass basis  coincide and 
\begin{eqnarray}
P_{ee}^M&\to &\overline{\rho_{11}} \nonumber \\
P_{ee}^{coh}&\to &\mid U_{11}\mid^2 \nonumber 
\end{eqnarray}
So 
\beq
P_{ee}^M(t)=P_{ee}^{coh}(t)+\int_0^t d\tau\ k(\tau)\rho^2(\tau)
P_{ee}^{coh}(t,\tau)P_{ee}^M(\tau)
\label{e9064}
\eeq
Eq.(\ref{e9064}) becomes exact for vanishing mixing angles.

A similar reasoning can be used to derive an equivalent expression for $P_{e\mu}^M$: in this case
$$P_{e\mu}^M=\trr \rho W $$
W is a certain matrix such that $W^2=0,\ W\cdot V=0$.
The final result is 
\beq
P_{e\mu}^M(t)=P_{e\mu}^{coh}(t)+\int_0^t d\tau\ k(\tau)\rho^2(\tau) P_{e\mu}^{coh}(t,\tau)P_{ee}^M(\tau)
\eeq
Defining total probabilities
$$P^{M,coh}=P_{ee}^{M,coh}+P_{e\mu}^{M,coh}$$
we obtain
\beq
P^M(t)=P^{coh}(t)+\int_0^t d\tau\ k(\tau)\rho^2(\tau)
P^{coh}(t,\tau)P_{ee}^M(\tau)
\label{e5066}
\eeq

This equation is specially useful to set an upper limit
 on $P^M$ or alternatively on the magnitude of the ''crossed'' terms that we
discarded.
 
We obtain the strict limit
\begin{eqnarray}
P^M(t)\leq P^{coh}(t)+\int_0^t k(t) \rho(t)
\end{eqnarray}

Setting $k(\tau)\leq k\exp \beta t$ ($\beta<2$ , eventually $\beta=0$) and
taking $t\to\infty$ it follows
 \begin{eqnarray}
P^M\leq P^{coh}+\frac{k\rho_0^2}{2-\beta}
\end{eqnarray}
In principle $P^M\equiv 1$, a departure from this
value signals a breakdown of the validity of the ''optical'' approximation.

If we denote by ''C'' the remaining ''crossed'' terms not incorporated in
the optical approximation then
\beq
1=P^M=P^{coh}+\left[\int\cdots \right]+C
\eeq
So
 \begin{eqnarray}
1-P^{coh}-\frac{k\rho_0^2}{2-\beta}\leq C
\end{eqnarray}

An actual estimate can be obtained if
the function $k(\tau)\rho^2(\tau)$ is of type exponential, so only the
values for $\tau\to 0$ of the integrand will contribute significantly. We take
the approximation
$$P^{coh}(t,\tau)P_{ee}^M(\tau)\simeq P^{coh}(t,0)
P_{ee}^M(\epsilon),$$
 with $\epsilon$ some small number.
Under this approximation:
\beq
P^M(t)=P^{coh}(t) \left (1+ k\rho_0^2 P_{ee}^M(\epsilon) \mu(t)\right )
\label{e7071}
\eeq

$P_{ee}^M(\epsilon)$ should be typically very small and not very dependent on
other parameters; 
Note that for $t\to\infty$ $\mu(t)$ only varies between $\approx 0.5-1$
irrespective of the exponential behavior of $k(t)$.

We see that the coherent density, or coherent probability, plays an important
role not only in its own right but also in the computation of the global density or probability.

\section{Computation of the Coherent probability.}
\label{s5}

The effective propagator defined in Eq.(\ref{e7465})
  is
\beq
\Sigma^{eff}(t,t_0)=T\exp \left( \frac{-i}{2}\right )\int_{t0}^t
dt' k(t')\rho(t')V_I(t') 
\label{e1005}
\eeq
So the effective potential becomes
\beq
V^{eff}_I(t)=\left( \frac{-i\rho(t)k(t)}{2}\right )V_I(t) 
\eeq
In the same case, optical propagator reads:
\begin{eqnarray}
\Sigma^{opt}(t,t_0)&=&1+\int_{t0}^t d\tau_1\int_{t0}^{\tau1} d\tau_2 \sVuno_{I}(\tau_1)
\Sigma^{opt}(\tau_1,\tau_2)\sVuno_I (\tau_2) \Sigma^{opt}(\tau_2,t_0)\nonumber \\
 &=& 1+\frac{-i }{2}\int_{t0}^t d\tau\ k(\tau)\rho(\tau)V_I(\tau) 
 \Sigma^{opt}(\tau,t_0)
\label{e9074}
\end{eqnarray}

The last relation in Eq.(\ref{e9074})
is just the integral equation for $\Sigma^{eff}$ which proves that both are identical in this case.

Here:
\beq
V^{eff}(t)=\frac{-i}{2}\rho_0^2 \exp(-2t)k(t) V
\label{e6071}
\eeq

For any function $k(t)$,
a possibility is to compute
 $\Sigma^{eff}$ by perturbation theory, considering the addition
 of
a ''small'' term  $V^{eff}(t)$ to the hamiltonian $H(t)$ of which we know
already  the exact solution.
In Section \ref{s6} we will follow this procedure for a particular $k(t)$. In
fact we will be able to sum the full perturbative series.

Another possibility, somehow less systematic, will be explored in the next
section where we will consider the small mixing limit.

\subsection{Small Mixing }
\label{s502}

Recalling the definition of $U_s$ in Eq.(\ref{e6003}) and inserting it in
Eq.(\ref{e6071}) we get
\beq
V_I^{eff}(t)=\frac{-\rho_0^2}{2}\exp(-2t)k(t)\pmatrix{\mid U_{s11}\mid^2
&U_{s11}U_{s12}^\ast\cr U_{s21}U_{s11}^\ast & \mid U_{s21}\mid^2}
\eeq
In the previous matrix we will further suppose we can neglect the off diagonal terms containing oscillating terms of the type
$\exp(iEt)$.
\beq
V_I^{eff}(t)\simeq\frac{-\rho_0^2}{2}\exp(-2t)k(t)
\pmatrix{\mid U_{r,11}\mid^2
&0\cr 0 & 1-\mid U_{r,11}\mid^2 }
\label{e3051}
\eeq
The time ordered integral is then trivial to perform, taking $t\to\infty$ we obtain
\beq
\Sigma^{eff}(\infty,t_0)=\pmatrix{\exp(-\frac{\rho^2_0}{2}\phi_1) 
& 0 \cr 
0 & \exp(\frac{\rho^2_0}{2}(\phi_1-\phi_2))} 
\eeq

where
\begin{eqnarray}
\phi_1(\rho_0,t_0)&=&\int_0^{\rho(t_0)/\rho_0}
 dx x k(\log(x)) \mid U_{r,11}(\rho_0 x)\mid^2\\
\phi_2&=&\int_0^{\rho(t_0)/\rho_0} dx x k(\log(x)) 
\end{eqnarray}

For $k(\log(x))\sim k x^\beta$ , both quantities $\phi_1,\phi_2$ are small, of
order $O(1)$.  For an arbitrary time t, the expressions will be analogous but
changing the inferior limit of integration into
 $\epsilon\approx 0$. So we expect
only a very slow dependence on t, this can be seen also from the presence of a
factor $\exp(-2t)$ in Eq.(\ref{e3051}).

In the small mixing limit  $\mid U_{r,11}\mid^2$ is the 
survival  probability
$P_{ee}$ in the absence of random perturbation (Eq.(\ref{e7007})). In a typical interesting
case (presence of resonance) $P_{ee}\equiv P_1\approx 1$,
for $\rho_0 e^{-t_0}>\rho_{res}$ 
and small ($P_0\approx 0$) for $\rho_0e^{-t_0}<\rho_{res}$. In the absence of resonance
(formally $\rho_{res}\to\infty$)
$P_{ee}\approx 1$ for all $\rho_0,t_0$. Using this behavior we can estimate
$\Sigma^{eff}(t\to\infty,t_0=0)$:
\begin{eqnarray}
\Sigma^{eff}(\rho_0>\rho_{res})&=& 
\pmatrix{ \exp \left [\frac{-\rho_{res}^2 k}{4}(P_1-P_0)-\frac{\rho_0^2
k}{4}P_0\right ] 
& 0 \cr 0 & \exp \left [\frac{\rho_{res}^2 k}{4}(P_1-P_0)+\frac{\rho_0^2
k}{4}(P_0-1)\right ] }\nonumber\\
& & \nonumber\\[0.5cm]
\Sigma^{eff}(\rho_0<\rho_{res})&=& 
\pmatrix{ \exp \frac{-\rho_0^2 k}{4}P_1
 & 0 \cr 0 & \exp \frac{-\rho_0^2 k}{4}(P_1-1)}
\label{e7078}
\end{eqnarray}

The coherent part of the macroscopic probability is (Eqs.(\ref{e4040}-\ref{e4041}))
\beq
P_{ee}^{M,coh}=\Sigma_i \left( \Sigma^{eff} \rho_{0I} \Sigma^{eff}\right ){}_{ii} V_{ii}
\eeq
For a $\Sigma^{eff}=
diag(A,B)$ as before, we have
\beq
P_{ee}^{M,coh}=A^2 \mid U_{r,11}^0\mid^2 V_{11}+B^2\mid U_{r,12}^0\mid^2 V_{22}
\eeq

\section{The special case k(t)=exp t.}
\label{s6}

\subsection{The Coherent Probability}
It is  
interesting to consider the case where $k(t)=k\exp t$. 
Then it is possible to further continue the exact analytical expressions.
 Eq.(\ref{e4056}) becomes:

\beq
<V_I(t) V_I(s)>= -i k \rho_0 \delta (t-s) V_I (t) 
\eeq
and  the effective potential is
\beq
V^{eff}_I=\left( \frac{-ik\rho_0}{2}\right )V_I ;\quad {\rm or\ }
V^{eff}(\rho_0)=V\left(\frac{-ik\rho_0^2}{2}\right )
\eeq
The effective potential has the same functional form as 
the original potential $\sim  \exp-t $ but with the constant $\rho_0$
redefined. $\Sigma^{eff}$ can be computed exactly in a simple
 way.New operators $U_s,U_r,U$ are defined by:
\beq
U_s^{0\dagger}(t)\Sigma^{eff}(t,t_0) U_s^0(t_0)=
U_s^{-1}(t)U_s(t_0)\equiv U(t,t_0)
\eeq
The new operators are just a renormalized version of the old ones
 $U_s^0,U^0$.
\beq
U(t,t_0;\rho_0)= U^0\left[t,t_0; \rho_0\ \left (1-\frac{i k \rho_0}{2}\right
)\right ]=\exp iH_0t\; U_r\left [\rho_0(1-\frac{ik\rho_0}{2})e^{-t}\right ]
\label{e6085}
\eeq

In the limit $t\to\infty$, both $U_s^0,U_s$, which  depend only on the
free hamiltonian $H_0$, are identical,
 so we get the  relation
\begin{eqnarray}
\Sigma^{eff}(t\to\infty,t_0)&=&U_s(t_0) U_s^{0\dagger}(t_0)\\
&=& U_r(\rho_{or}) U_r^\dagger(\rho_0) \quad (t_0=0)
\end{eqnarray}

The asymptotic density (Eqs.(\ref{e4029}-\ref{e4037})) is given by
\beq
\rho_{as}^{coh}=U_s \rho_0 U_s^\dagger
\eeq
The  coherent probability can be computed using 
Eq.(\ref{e4040}) which amounts essentially to redefine $\rho_0$ in Eq.(\ref{e7006}).

The total incoherent part  
\begin{eqnarray}
P^{M,in}=1-\left (\mid U_{r,11}\mid^2+\mid U_{r,12}\mid^2\right )
\end{eqnarray}
 is not equal to zero because the matrices $U_s,U_r$ are not unitary now.

\section{Some numerical results}

We see that the  choice
 $k(t)\sim k_1\exp t$ allows for a fully analytical
exact computation of the coherent probability. There are 
no physical grounds for
this choice, as there are no physical grounds for any other selection, for
example $k(t)\sim k_2$), as long as
we don't have a very detailed knowledge of the Sun structure. If we suppose that
random fluctuations are only important if they happen in the resonance layer
(but see the warning commentary in a previous section) 
the result of both choices should be  approximately equivalent
if we take $k_1,k_2$ such that
\begin{eqnarray}
k_1 \exp t_{res}&\approx& k_2\\
{\hbox{or}}\quad\quad\quad\quad  k_1\rho_0&\approx& k_2\rho_{res}
\end{eqnarray}
 where $\rho_{res},t_{res}$ are the position and local density of the resonance layer.

Outside the resonance region, for $t\to\infty$,  $k_1\exp t >> k_2$ but in anycase
the products $k_1\exp t\rho^2(t),k_2\rho^2(t)\to 0$.

For $t_{res}>t\to 0$, $k_1 \exp t=\exp(t-t_{res}) k_2$; for a distance
$(t-t_{res})$ equivalent to a quarter of the solar radius that means $k_1\exp
t\approx k_2/3$, 
for half of the solar radius $k_1\exp
t\approx k_2/10$. 
The differences between both cases are not so strong as it could be thought due
to the presence of the extra exponential.

\begin{figure}[p]
\centering\hspace{0.8cm}
\begin{tabular}{c}
{\epsfig{file=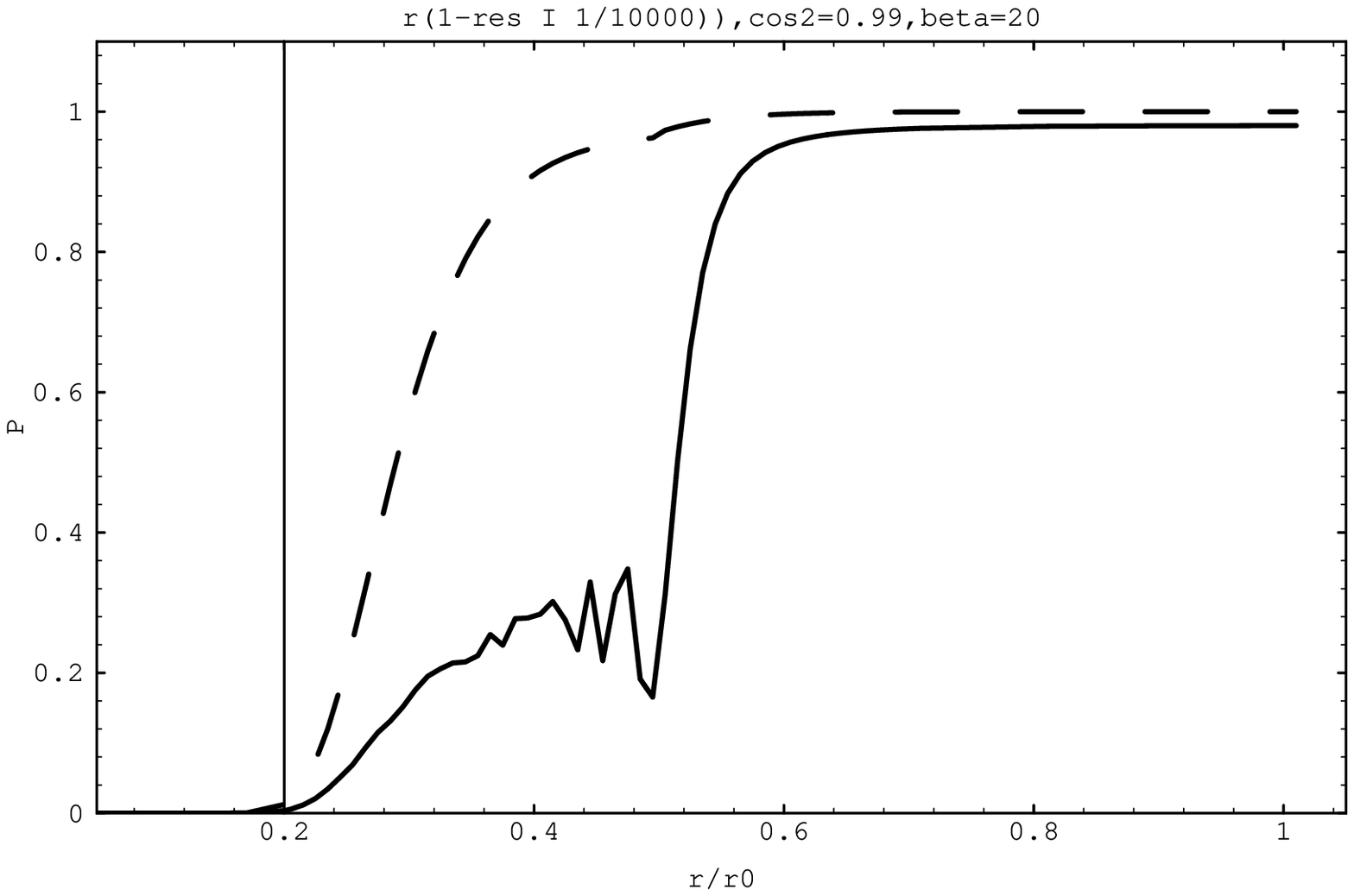,height=10cm}}\\[-4cm]
{\epsfig{file=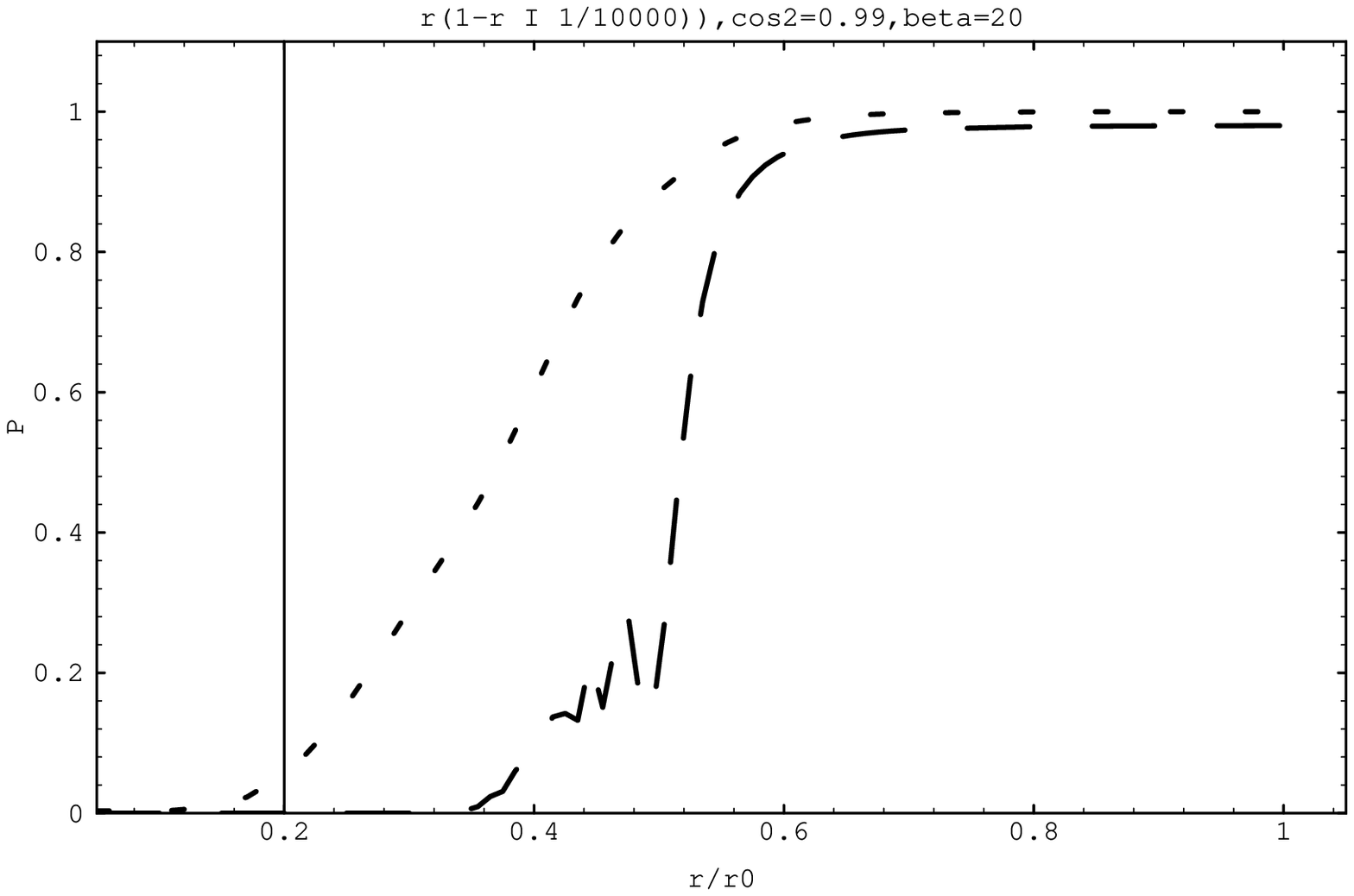,height=10cm}}\\[-1cm]
\end{tabular}
\caption{The $\nu_e$ coherent survival probability as function of the neutrino creation
point. 
(Top) $P_{ee}^{coh}$ (continuos) and $P^{coh}$ (dashed) computed using
$\Sigma^{eff}$ given by Eq.(\protect\ref{e7078}).
Bottom, the same probabilities computed  using
$U_r(\rho_0(1-ik\rho_0/2))$. ($k=10^{-4},\beta=20,\cos^2\theta=0.99$ for both figures.)}
\label{f3}
\end{figure}

In Fig.(\ref{f3}) we plot the averaged coherent survival probability $P_{ee}^{coh}$
and the total coherent probability $P^{coh}$.
In the top figure we
make use of the Expression
(\ref{e7078}) for $\Sigma^{eff}$.
In the bottom one we plot  the raw Formula (\ref{e6085}). 
In both figures 
the probability is
strongly suppressed for neutrinos created near the origin.
In the bottom one,
the equivalent fluctuation level at the resonance region varies for 
different creation point $r/r_0$, but, in spite of the exponential
behavior of $k(t)$, this variation is rather modest.
For $k=10^{-4}$ as used, 
 the equivalent fluctuation  at the resonance region
$r/r_0\approx 0.5-0.6$ is only $\approx 7\%$ of the local density for a neutrino created at $r/r_0=1/10$
and $\approx 13\%$ for a neutrino created at the center.

\begin{figure}[p]
\centering\hspace{0.8cm}
{\epsfig{file=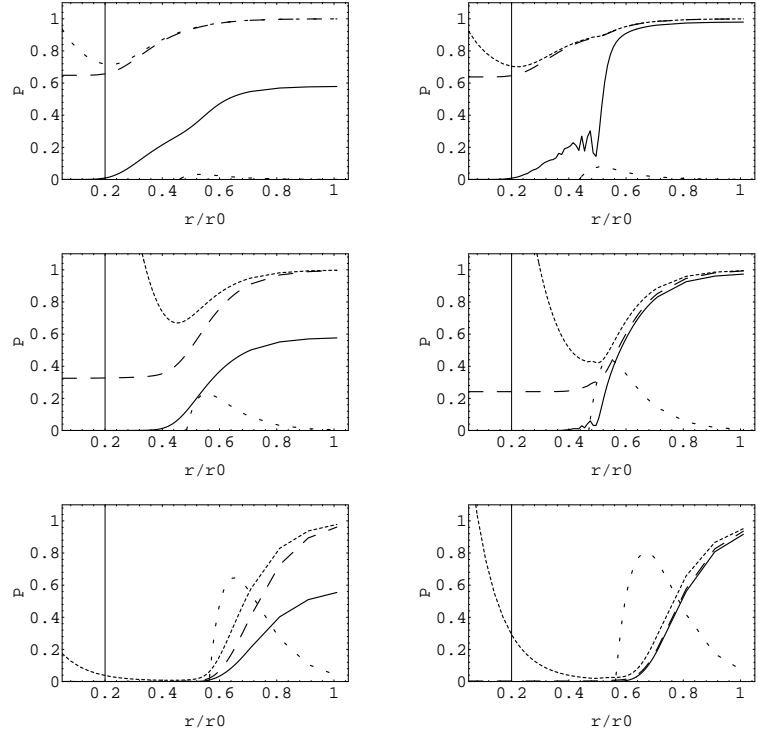,height=15cm}}
\caption{Coherent probabilities $P^{coh}$ (longer dash), $P_{ee}^{coh}$
(continuous line) 
as a function of $k$ and the mixing angle. They are computed using
$U_r(\rho_0(1-ik\rho_{res}/2))$. $k=10^{-4},10^{-3},10^{-2}$ for upper, middle,
lower figures respectively. Right figures: large mixing, $\cos^2\theta=0.70$;
left, small mixing $\cos^2\theta=0.99$. In all the cases 
$\beta=20$.
It is also depicted $P^M$ (Eq.(\protect\ref{e7071})) (short denser dash)  and
$C_{min}=1-P^{coh}-k\rho^2/2$ (short less denser dash).}
\label{f4}
\end{figure}

In Fig.(\ref{f4}) we plot instead the averaged coherent probabilities computed
using the corrected formula
$\rho_{or}=\rho_0(1-ik\rho_{res}/2)$ which guarantees at least uniformity at the
resonance region for different values of $k$.
 
For small mixing angles,
$P_{ee}^{coh}$  shows a moderate damping as $k$ increases.
 The variation for the
total coherent probability is much  stronger. Beyond the resonance
region, practically all $P^{coh}\approx P_{ee}^{coh}$. 

As the mixing angle increases
the pattern is different, $P^{coh}$ practically
doesn't suffer alteration but
$P_{ee}^{coh}$ is strongly diminished.

For both angles:
for $k=10^{-3},10^{-2}$ , $P^{coh}$
 is practically zero in
 the inner creation regions. Here
nearly all the eventual survival probability must come
 from the diffuse, incoherent
scattering.  

It is also shown
the $P^M$ defined by Eq.(\ref{e7071}). For illustration purposes
we choose the maximal value  $P_{ee}^M(\epsilon)=1$. Even in this maximal case
$P^M$ is far from unity,
this is particularly evident for the strongest fluctuations.
For large regions the ''optical'' approximation hardly differs from the coherent
probability and fails to be an appreciable improvement.
The lower limit for the ''crossed'' or 
''non-optical'' corrections , although without  much value numerically, can be interpreted as
indicative that it 
is precisely around the resonance region where the ''optical''
approximation failure is strongest.

\begin{figure}[p]
\centering\hspace{0.8cm}
\begin{tabular}{cc}
{\epsfig{file=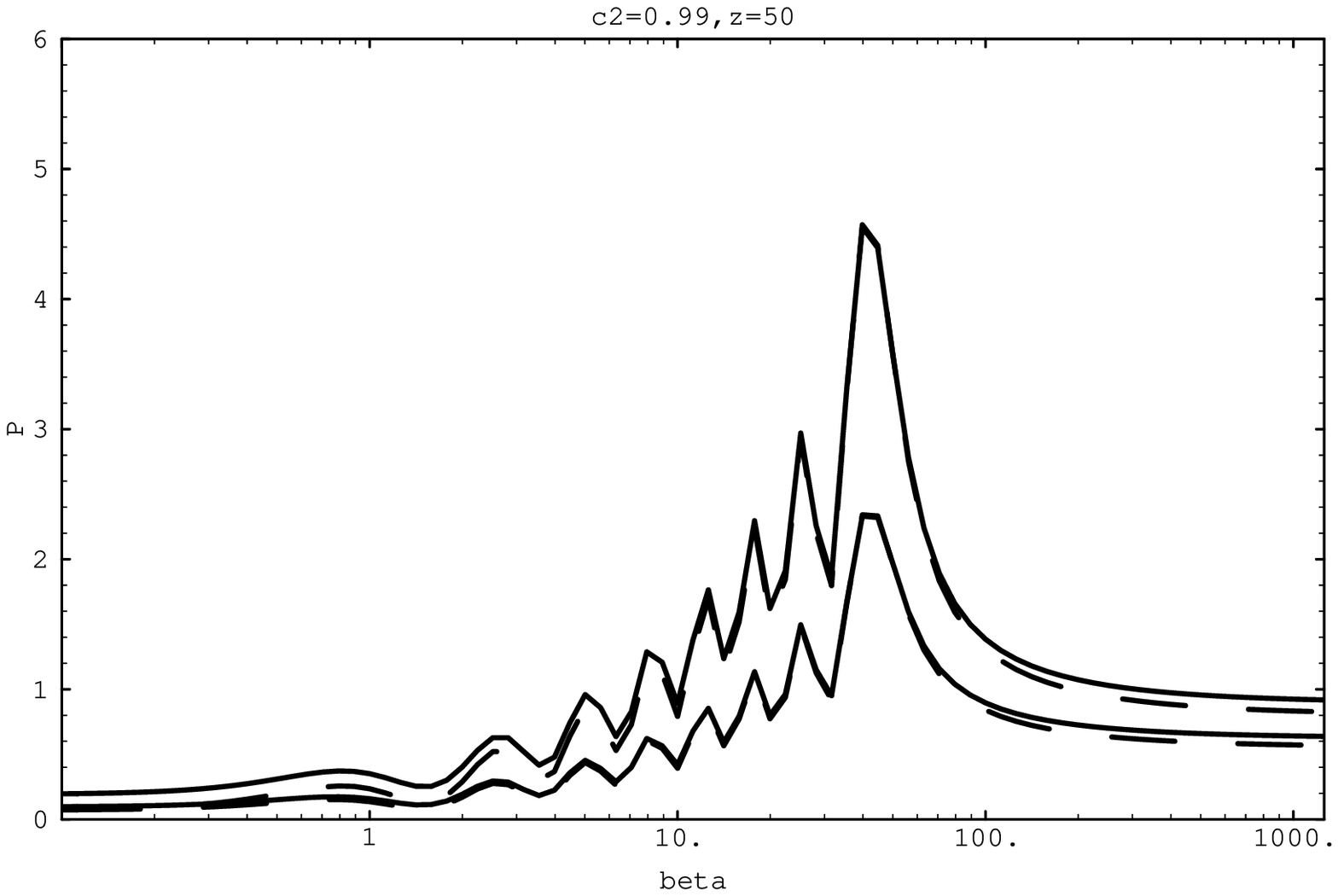,height=6cm}}
 &
{\epsfig{file=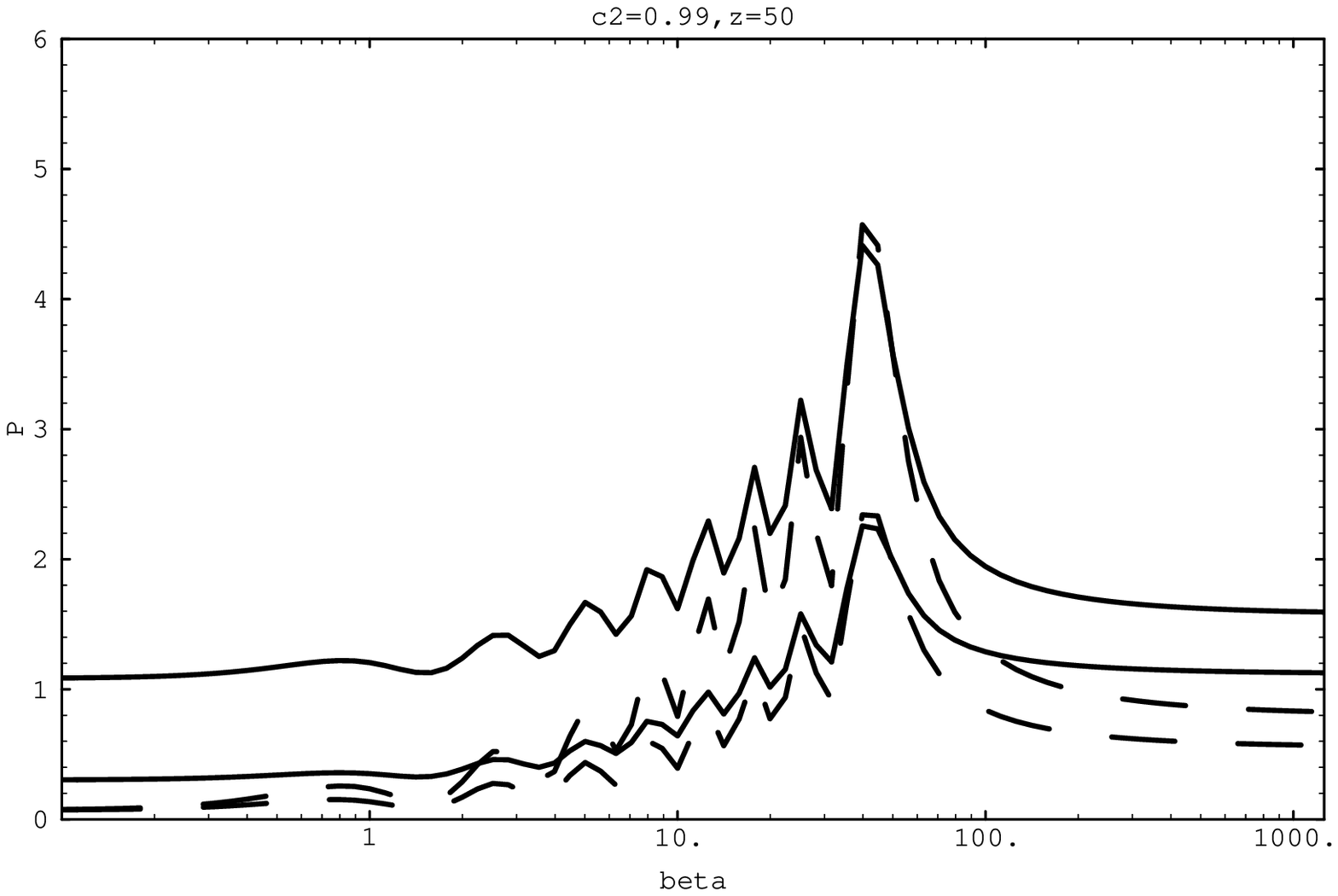,height=6cm}}
\\[-1cm]
{\epsfig{file=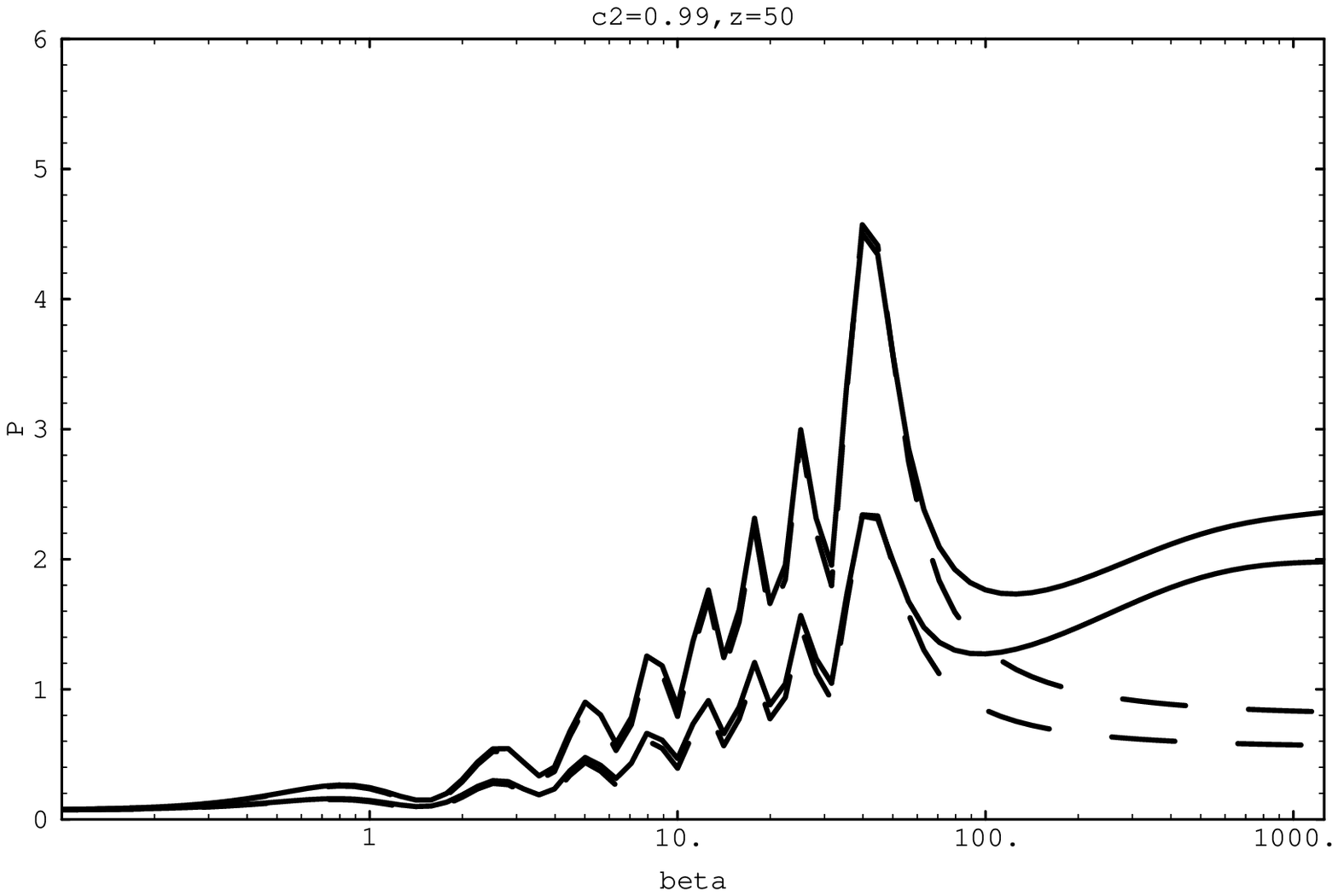,height=6cm}}&
{\epsfig{file=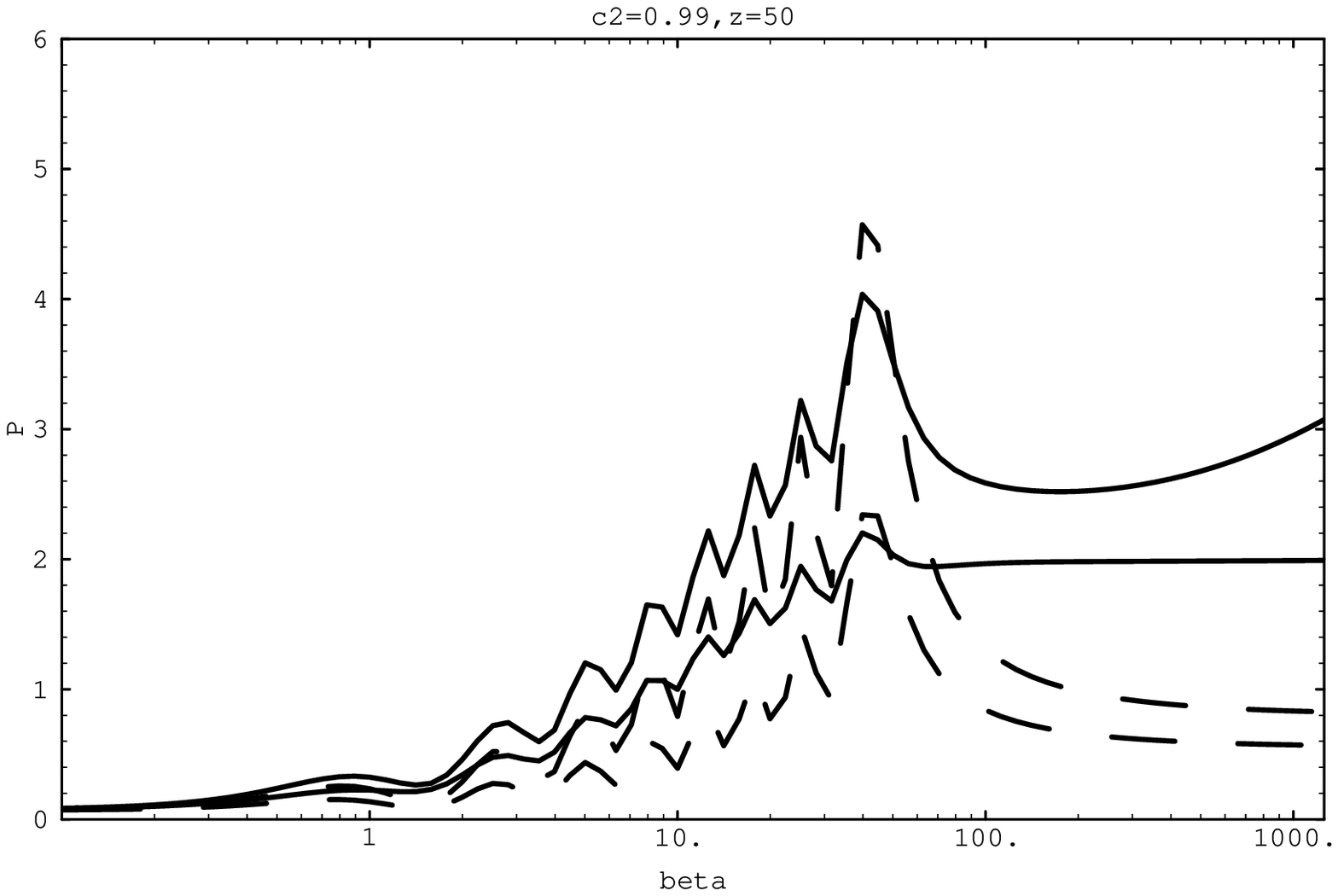,height=6cm}}\vspace{-1cm}
\end{tabular}
\caption{ $\tWW\sigma_1^1$ (line) and
$\sigma_2$ (dashed line) for a neutrino with the same characteristics as in
Fig.(\protect{\ref{f1}}) in presence of a random perturbation. 
Two top figures: average amplitude computed using $\Sigma^{eff}$ from 
Eq.(\protect\ref{e7078}). Lower figures: using $U_r(\rho_{or})$.
 $k=10^{-4},10^{-3}$ ($\approx 1\%, 5\%$ fluctuations ) 
respectively left and right.}
\label{f5}
\end{figure}

The average of the cross sections defined by Eq.(\ref{e6024}-\ref{e6025}) is
obtained inserting the average value of the 
matrix U, $U^{eff}$, computed using 
either
$\Sigma^{eff}$ from Eq.(\ref{e7078}) (Fig.(\ref{f5}), top)
or $U_r(\rho_{or})$ (the two bottom figures). In both cases 
 the approximated expression 
$\rho_{res}\simeq \beta \cos 2\theta$ has been used.
For the smaller   $k\approx 10^{-4}$ (or $1\%$ 
fluctuations) the effect of the
random perturbation is nearly  negligible in both cases. The discrepancy in the
large $\beta$ behavior has the same origin as the difference between
Fig.(\ref{f3}) (top) and Fig.(\ref{f4}). 
 For $k\approx 10^{-3},10^{-2}$ the effect can be appreciable. 
The interpretation of these plots in terms of concrete survival probabilities is
problematic and subject to further study, 
but what it is clear already from them is
that  the
basic behavior, the very existence, position and 
width of a peak corresponding to the resonance layer
remains
 unaltered. There are no discrepancies in this region between the two approaches.

\begin{figure}[p]
\centering\hspace{0.8cm}
\begin{tabular}{c}
{\epsfig{file=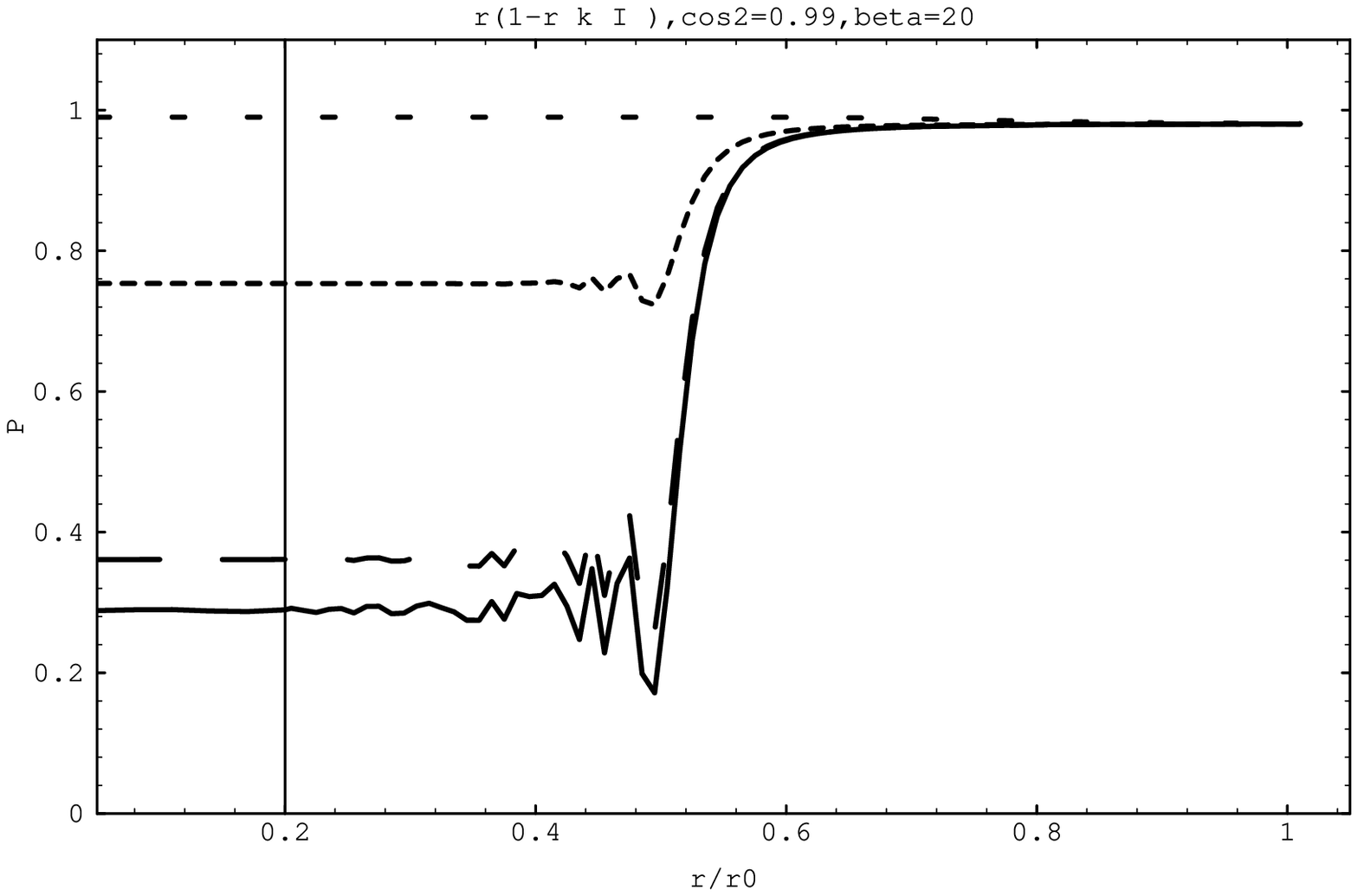,height=10cm}}\\[-1cm]
{\epsfig{file=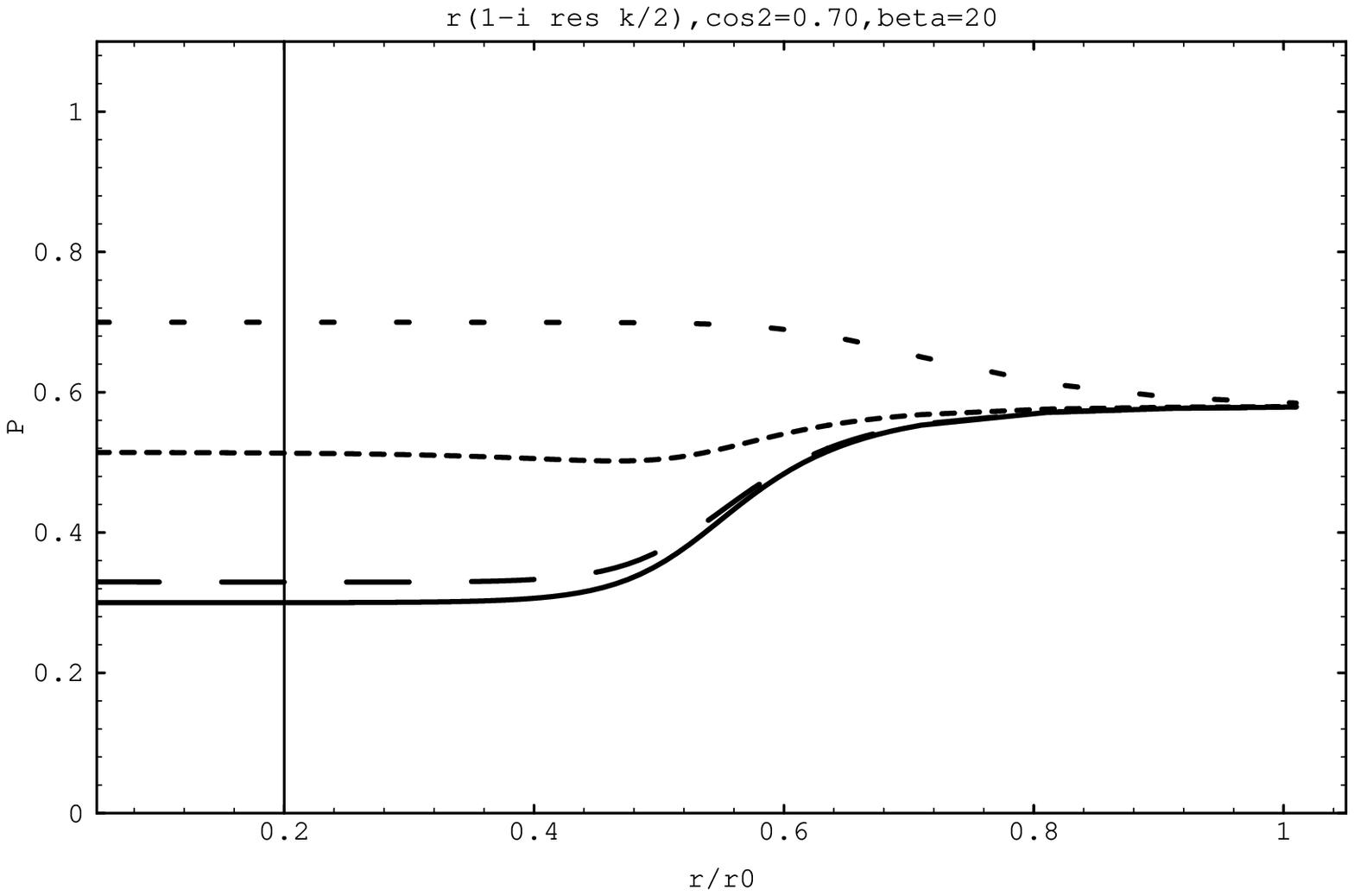,height=10cm}}\\[-1cm]
\end{tabular}
\caption{The $\nu_e$ survival probability as function of the neutrino creation
point computed using the {\em ansatz} of Section \protect\ref{s22},
Eq.(\protect\ref{e7006}) with $\rho_{or}=\rho_0(1-ik\rho_{res}/2)$. 
Continuos line: non-random probability ($k=0$).
Dashed lines: $k=10^{-4},10^{-3},10^{-2}$ respectively.
For both figures $\beta=20$;  $\cos^2\theta=0.99$ (Top), $\cos^2\theta=0.70$ (Bottom).}
\label{f6}
\end{figure}

In Fig.(\ref{f6}), we plot the averaged survival probability 
for different $k$ using
the {\em ansatz} formula for $P_{ee}(\rho_{or})$ of Section
\ref{s22} and $\rho_{or}=\rho_0(1-i \rho_{res}k/2$.
There is consistency between the behavior expressed here and that of $P_{ee}^{coh},P^{coh}$.
If we take at face-value this plot and we compare it with
Fig.(\ref{f3}) we arrive at the conclusion that 
for neutrinos created much before the resonate region the
total survival probability comes  essentially from the 
large avalaible quantity of ''incoherent'' probability.
As happens in other physical circumstances, the presence
of ''diffuse''
scattering disfavors transitions $\nu_e\to\nu_{\mu}$ and produces an
enhancement of $P_{ee}$ (state localization). 
In the same figure, 
the quick increase 
of the effect of the random perturbations with $k$ 
is somehow surprising; for $\approx 10\%$
fluctuations the effect is already quite considerable (''reverse'' MSW effect in
the case of large mixing angle).

\section{Brief summary of results and further conclusions.}

The most important result of this work is the
derivation of analytical exact expressions for the average {\em coherent}
transition probability for a special case of colored $\delta$-correlated Gaussian
noise. We have shown that in this case the consequence of the presence of the
noise is a complex renormalization of the matter density.
Other approximative expressions for more general cases have
 been  developed as well.
It has been suggested 
an enhancement of the survival probability 
when the {\em
incoherent} probability becomes dominant.

It has been proposed for the first time
to consider  new scattering ''cross sections''. It has
been shown how they are able to describe themselves the MSW effect; clearly   further work
has to be done in this respect. The main importance of these
quantities is that their statistical average is computable in a simple way.

The general conclusion is that the MSW effect survives the presence of random
perturbations at least for small
 fluctuation values. There are indications
that their  effect can become important for large fluctuations.

\vspace{1cm}
{\Large{\bf Acknowledgments.}}\vspace{0.3cm}

I would like to thank to Peter Minkowski for many enlightening discussions. This 
work has been supported in part by the Wolferman-Nageli Foundation (Switzerland)
and by the MEC-CYCIT (Spain).

\newpage

\newpage

\begin{thebibliography}{10}

\bibitem{wol1}
{ L.~Wolfenstein}, {\em Neutrino oscillation in matter}, Phys. Rev. D., 17
  (1978), pp.~2369--2376.

\bibitem{mik1}
{  S.P.~Mikheyev, A.Y. Smirnov}, {\em Resonant amplification of neutrino oscillations in
  matter and solar neutrino spectroscopy}, Il Nuovo Cimento, 9C (1986),
  pp.~17--26.

\bibitem{ber1}
{ G.~Bergmann}, {\em Physical interpretation of weak localization: A
  time-of-flight experiment with conduction electrons}, Phys. Rev. D., 28
  (1983), pp.~2914--2920.

\bibitem{vol1}
{  D.~Vollhardt, P. Wolfle}, {\em Diagrammatic, self-consistent treatment of the
  anderson localization problem in d<= 2 dimensions}, Phys. Rev. B., 22 (1980),
  pp.~4666--4679.

\bibitem{fitz1}
{ R.M.~Fitzgerald, A.A.~Maradudin, F. Pincemin}, {\em Scattering of a scalar wave from a
  two-dimensional randomly rough neumann surface}, Waves Random Media, 22
  (1995), pp.~381--402.

\bibitem{lor1}
{ F.N.~Loreti, A.B. Balantekin}, {\em Neutrino oscillations in noisy media}, Phys. Rev. D.,
  50 (1994), pp.~4762--4770.

\bibitem{ane1}
{  C.~Aneziris, J. Schechter}, {\em Neutrino spin-rotation in a twisting magnetic field},
  SU-4228-449.

\bibitem{akh1}
{ E.K.~Akhmedov, S.T. Petcov}, {\em Neutrinos with mixing in twisting magnetic fields},
  Phys. Rev. D., 48 (1993), pp.~2167--2170.

\bibitem{nic2}
{ A.~Nicoladis}, 
{\em Random magnetic fields in the sun and solar neutrinos},
Phys. Lett. 262 (1991) 2,3, pp.~303-306.


\bibitem{lor2}
{  F.N.~Loreti, YZ~Qian, G.M. Fuller, A.B. Balantekin}, {\em Effects of random density fluctuations
  on matter-enhanced flavor transitions in supernovae and implications for
  supernove dynamics and nucleosynthesis}, astro-ph/9508106,  (1995).

\bibitem{nun2}
{  H.~Nunokawa, A.~Rossi, V.B. Semikoz, J.W.F. Valle}, {\em The effect of random matter density
  perturbations on the msw solution to the solar neutrino problem}. IFIC/95-49, hep-ph/9602307.




\bibitem{emi1}
{ E.~Torrente-Lujan}, {\em Exact analytic description of neutrino oscillations
  in matter with exponentially varying density for an arbitrary number of
  neutrino species}, Phys. Rev. D., 53 (1996), pp.~53--67.





\bibitem{aga1}
{ D.~Aggasi, C.M.~Ko, H.A. Weidenmuller}, {\em Transport theory of deep inelastic heavy-ion
  collisions based on a random-matrix model. i derivation of the trasport
  equation}, Ann. of Phys., 107 (1977), pp.~140--167.



\end{thebibliography}

\end{document}